\documentclass[aps,pra,twocolumn]{revtex4-1}
\usepackage{amsmath,amsfonts,amssymb}
\usepackage[caption=false]{subfig}
\usepackage{tikz}
\usetikzlibrary{shapes}

\newcommand{\ket}[1]{\ensuremath{|#1\rangle}}
\newcommand{\bra}[1]{\ensuremath{\langle#1|}}
\newcommand{\ketbra}[2]{\ensuremath{\ket{#1}\!\bra{#2}}}

\newcommand{\In}{\ensuremath{\textrm{in}}}
\newcommand{\Out}{\ensuremath{\textrm{out}}}

\tikzstyle{vtx}=[circle, draw, fill=black,
                 minimum size=3pt, inner sep=0pt]
\tikzstyle{edge}=[thick]
\tikzstyle{tailv}=[circle, draw, fill=black, minimum size=2pt, inner sep=0pt]
\tikzstyle{tail}=[dashed]

\newcommand{\lgndSquareFilled}[3]{
  \draw[line width=0.1pt,fill=black] (#1-#3/2,#2-#3/2)
    --++(#3,0) --++(0,#3) --++(-#3,0) -- cycle;
}
\newcommand{\lgndTriDownFilled}[3]{
  \draw[line width=0.1pt,fill=black] (#1-#3/2,#2+#3/2)
    --++(#3,0) --++(-#3/2,-#3) -- cycle;
}
\newcommand{\lgndTriUpFilled}[3]{
  \draw[line width=0.1pt,fill=black] (#1-#3/2,#2-#3/2)
    --++(#3,0) --++(-#3/2,#3) -- cycle;
}
\newcommand{\lgndTriDown}[3]{
  \draw[line width=0.1pt] (#1-#3/2,#2+#3/2)
    --++(#3,0) --++(-#3/2,-#3) -- cycle;
}
\newcommand{\lgndPlus}[3]{
  \draw[line width=0.1pt] (#1,#2)
    --++(-#3/2,0) --++(#3,0) --++(-#3/2,0)
    --++(0,#3/2) --++(0,-#3);
}
\newcommand{\lgndCircle}[3]{
  \draw[line width=0.1pt] (#1,#2) circle (#3/2);
}
\newcommand{\lgndTimes}[3]{
  \draw[line width=0.1pt] (#1,#2)
    --++(-#3/2,-#3/2) --++(#3,#3) --++(-#3/2,-#3/2)
    --++(#3/2,-#3/2) --++(-#3,#3);
}
\newcommand{\lgndTriUp}[3]{
  \draw[line width=0.1pt] (#1-#3/2,#2-#3/2)
    --++(#3,0) --++(-#3/2,#3) -- cycle;
}
\newcommand{\lgndSquare}[3]{
  \draw[line width=0.1pt] (#1-#3/2,#2-#3/2)
    --++(#3,0) --++(0,#3) --++(-#3,0) -- cycle;
}
\newcommand{\lgndBowtie}[3]{
  \draw[line width=0.1pt] (#1,#2)
    --++(#3*0.92388*1.45/2, #3*0.382683*1.45/2)
    --++(-#3*0.92388*1.45, -#3*0.382683*1.45)
    --++(#3*0.92388*1.45/2, #3*0.382683*1.45/2)
    --++(-#3*0.92388*1.45/2, #3*0.382683*1.45/2)
    --++(#3*0.92388*1.45, -#3*0.382683*1.45);
}
\newcommand{\lgndTeepee}[3]{
  \draw[line width=0.1pt] (#1,#2)
    --++(#3*0.382683*1.45/2, #3*0.92388*1.45/2)
    --++(-#3*0.382683*1.45, -#3*0.92388*1.45)
    --++(#3*0.382683*1.45/2, #3*0.92388*1.45/2)
    --++(#3*0.382683*1.45/2, -#3*0.92388*1.45/2)
    --++(-#3*0.382683*1.45, #3*0.92388*1.45);
}
\newcommand{\lgndDiamond}[3]{
  \draw[line width=0.1pt] (#1-#3/2,#2)
    --++(#3/2,#3/2) --++(#3/2,-#3/2) --++(-#3/2,-#3/2) -- cycle;
}
\def\vtxR{1.75pt}
\newcommand{\vtxIn}[1]{
  \draw[fill=black]
    (#1) --++(0,-\vtxR) arc (-90:90:\vtxR) -- cycle;
  \draw[fill=white,color=white,,line width=0.2pt]
    (#1) --++(0,\vtxR) arc (90:270:\vtxR) -- cycle;
  \draw[thick] (#1) circle (\vtxR);
}
\newcommand{\vtxOut}[1]{
  \draw[fill=black]
    (#1) --++(0,\vtxR) arc (90:270:\vtxR) -- cycle;
  \draw[fill=white,color=white,line width=0.2pt]
    (#1) --++(0,-\vtxR) arc (-90:90:\vtxR) -- cycle;
  \draw[thick] (#1) circle (\vtxR);
}
\newcommand{\vtxIO}[1]{
  \draw[thick,fill=white] (#1) circle (\vtxR);
}
\newcommand{\vtx}[1]{
  \draw[fill=black] (#1) circle (\vtxR);
}

\begin{document}

\title{Single-qubit unitary gates by graph scattering}
\author{Benjamin A.\ Blumer}
\thanks{These authors contributed equally to this work.}
\author{Michael S.\ Underwood}
\thanks{These authors contributed equally to this work.}
\author{David L.\ Feder}
\email[Corresponding author: ]{dfeder@ucalgary.ca}
\affiliation{Institute for Quantum Information Science,
University of Calgary, Alberta T2N 1N4, Canada}

\date{November 17, 2011}

\begin{abstract}
We consider the effects of plane-wave states scattering off finite graphs, as
an approach to implementing single-qubit unitary operations within the 
continuous-time quantum walk framework of universal quantum computation. 
Four semi-infinite tails are attached at arbitrary points of a given graph, 
representing the input and output registers of a single qubit. For a range
of momentum eigenstates, we enumerate all of the graphs with up to $n=9$ 
vertices for which the scattering implements a single-qubit gate.
As $n$ increases,
the number of new unitary operations increases exponentially, and for $n>6$
the majority correspond to rotations about axes distributed roughly uniformly
across the Bloch sphere. Rotations by both rational and irrational multiples
of $\pi$ are found.
\end{abstract}

\maketitle

\section{Introduction}

Quantum walks are the quantum mechanical analogs of random walks in classical
systems, and like their classical counterparts they have provided a useful 
framework for the construction of efficient quantum 
algorithms~\cite{Ambainis:2003,Kempe:2003,Kendon:2007,Kendon:2011}.
In addition 
to the development of quantum walk approaches to known quantum algorithms such 
as the Grover search~\cite{Santha:2008}, triangle finding~\cite{Magniez:2005}, 
and element distinctness~\cite{Ambainis:2007}, several quantum algorithms were 
obtained using quantum walks that were previously not found within the 
standard quantum circuit or adversary models, such as the traversal of 
glued-tree graphs with randomness~\cite{Childs:2003}, and the solution of NAND 
trees~\cite{Farhi:2008, Childs:2009a}, min-max trees~\cite{Cleve:2008}, and 
general boolean functions~\cite{Reichardt:2008,Ambainis:2010}. More recently it
has been shown that quantum walks are able to perform arbitrary quantum
computational tasks~\cite{Childs:2009b,Lovett:2010,Underwood:2010}.

In the continuous-time approach to universal quantum walks~\cite{Childs:2009b},
computational basis states are represented by long linear graphs 
(tails) which support plane wave modes characterized by momentum eigenstates. 
A single quantum walker is initialized in a momentum eigenstate of a single
semi-infinite line out of $2^N$ such tails representing $N$ qubits. Quantum
operations on the encoded information are carried out by interspersing the 
tails with small graphs. The scattering of the plane waves off these graphs
mixes the amplitudes on different tails and adds possible phase shifts. 

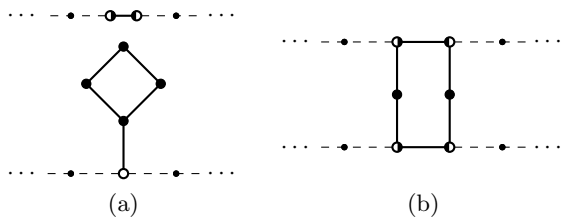
\begin{figure}[t]
  \begin{center}
    \subfloat{
      \label{subfig:Childsgraphs-PhaseShift}
      \tikzset{x=0.7cm,y=0.7cm}
      \begin{tikzpicture}
        \node[vtx] (v1) at (0,0) {};
        \node[vtx](v2) at (0,1) {};
        \node[vtx](v3) at (0.707,1.707) {};
        \node[vtx](v4) at (0,2.414) {};
        \node[vtx](v5) at (-0.707,1.707) {};
        \vtxIO{v1}
        \vtx{v2}
        \vtx{v3}
        \vtx{v4}
        \vtx{v5}
        \node[vtx] (v6) at (-0.25,3) {};
        \node[vtx] (v7) at (0.25,3) {};
        \vtxIn{v6}
        \vtxOut{v7}
        \draw[edge] (v1) -- (v2) -- (v3) -- (v4) -- (v5) -- (v2);
        \draw[edge] (v6) -- (v7);
        \node at (0,-0.6) {\subref{subfig:Childsgraphs-PhaseShift}};
        %
        \foreach \x in {1,1} {
          \node[tailv] at (-\x,3) {};
          \node[tailv] at (-\x,0) {};
          \node[tailv] at (\x,0) {};
          \node[tailv] at (\x,3) {};
        };
        \draw[tail] (v1) -- (-1.5,0);
        \draw[tail] (v1) -- (1.5,0);
        \draw[tail] (v6) -- (-1.5,3);
        \draw[tail] (v7) -- (1.5,3);
        \node at (-1.9,0) {$\cdots$};
        \node at (-1.9,3) {$\cdots$};
        \node at (1.9,0) {$\cdots$};
        \node at (1.9,3) {$\cdots$};
      \end{tikzpicture}
    }
    \subfloat{\label{subfig:Childsgraphs-BasisChanger}
    		\tikzset{x=0.7cm,y=0.7cm}
      \begin{tikzpicture}
        \node[vtx] (v1) at (0,0) {};
        \node[vtx] (v2) at (1,0) {};
        \node[vtx](v3) at (1,1) {};
        \node[vtx] (v4) at (1,2) {};
        \node[vtx] (v5) at (0,2) {};
        \node[vtx](v6) at (0,1) {};
        \vtxIn{v1}
        \vtxOut{v2}
        \vtx{v3}
        \vtxOut{v4}
        \vtxIn{v5}
        \vtx{v6}
        \draw[edge] (v1) -- (v2) -- (v3) -- (v4) -- (v5) -- (v6) -- (v1);
        \node at (0.5,-1.1) {\subref{subfig:Childsgraphs-BasisChanger}};
        %
        \foreach \x in {1,1} {
          \node[tailv] at (-\x,0) {};
          \node[tailv] at (-\x,2) {};
          \node[tailv] at (\x+1,0) {};
          \node[tailv] at (\x+1,2) {};
        };
        \draw[tail] (v1) -- (-1.5,0);
        \draw[tail] (v2) -- (2.5,0);
        \draw[tail] (v5) -- (-1.5,2);
        \draw[tail] (v4) -- (2.5,2);
        \node at (-1.9,0) {$\cdots$};
        \node at (-1.9,2) {$\cdots$};
        \node at (2.9,0) {$\cdots$};
        \node at (2.9,2) {$\cdots$};
      \end{tikzpicture}
    }
  \end{center}
\caption{\label{fig:Childsgraphs}
  The graphs identified in Ref.~\cite{Childs:2009b} that generate a 
universal set of one-qubit unitary gates, with vertices denoted by circles
and edges by solid lines. Open circles indicate that both an input and
an output tail attach at that vertex, whereas circles open only to the left
or right indicate attachment of only input or output tails, respectively.
Graphs \subref{subfig:Childsgraphs-PhaseShift} and
\subref{subfig:Childsgraphs-BasisChanger} generate the rotations $R_Z(\pi/4)$ 
and $R_X(\pi/2)$, respectively, for $k=-\pi/4$. The tails, depicted by dashed
edges and ellipses to denote their continuation, are included to indicate the
attachment vertices.
These graphs appear in the supplemental material as IDs 238 and 309.
}
\end{figure}

In Ref.~\cite{Childs:2009b}, a small number of such graphs were identified 
that could generate a universal set of $N$-qubit unitary gates for a fixed 
momentum $k=-\pi/4$ (all length units are suppressed for convenience). The two 
graphs needed for single-qubit operations are shown in 
Fig.~\ref{fig:Childsgraphs}. The first of these, depicted in 
Fig.~\ref{subfig:Childsgraphs-PhaseShift}, is a disconnected
seven-vertex graph that
yields the unitary $R_Z(-\pi/4)=\mbox{diag}(1,e^{-i\pi/4})\equiv T$ neglecting 
unimportant overall phases ($R_a(\theta)$ denotes a rotation about axis $a$ by 
an angle $\theta$, and $X$, $Y$, and $Z$ are the Pauli matrices). The upper 
subgraph is a two-site graph, so with the tails encoding the 
$\ket{1}_{\rm input}$ and $\ket{1}_{\rm output}$ attached it has the effect of
changing the length of the full tail by one lattice spacing. For plane waves 
of the form $e^{ikx}$, this corresponds to a scattering phase shift of 
$-\pi/4$ on the $\ket{1}$ state. The lower subgraph, attached to the $\ket{0}$ 
input and output tails on the same vertex, produces no scattering phase shift 
and one thus obtains the desired rotation. This subgraph furthermore has the 
desirable property of also having an {\it effective length} of one lattice 
spacing, so that the registers $\ket{0}$ and $\ket{1}$ maintain their spatial 
coherence after the scattering event.  Under this scheme
the various paths through a graph be of equal length, and there must 
be an identity gate with the same effective length for a given momentum to
act on all registers where no gate is desired. This identity requirement is
trivial to satisfy in the case of graphs with integral effective lengths, but 
as we show below graphs can have lengths that are non-integral, or even 
irrational.

The second graph, shown in 
Fig.~\ref{subfig:Childsgraphs-BasisChanger},
is a connected graph that yields $R_X(\pi/2)
\equiv V$, again neglecting unimportant overall phases. The Hadamard gate can 
be constructed via $H=T^2VT^2$, and combinations of $H$ and $T$ can produce 
any single-qubit unitary at fixed accuracy~\cite{Kaye:2007}.
While these two graphs are sufficient to generate a universal set of 
single-qubit gates, and a universal set of gates when an entangling gate is 
included~\cite{Childs:2009b}, it is not clear if there are other graphs that 
could be chosen instead.

An immediate question is: are there other graph combinations 
that work equally well for $k=-\pi/4$, or for other values of the momentum? 
Because the number of non-isomorphic graphs grows exponentially in the number 
of vertices $n$, one might expect a plethora of choices which would be useful
in the design of quantum algorithms based on quantum walks. Yet it is also 
conceivable that larger graphs will simply reproduce the unitaries found for
smaller graphs, or that as $n$ increases the requirement that there be
perfect transmission from each of the input tails to the combination of the
two outputs, with equal effective lengths on all four paths,
will become increasingly difficult to satisfy.

In the present work, we explore these issues numerically by calculating the 
single-qubit unitary operators resulting from two input tails scattering
to two output tails, off every 
possible graph with up to nine vertices. Momentum eigenstates $k=(p/q)\pi$ 
where $q\in\{2,3,4,5\}$ and $0<p<q$ for integer $p$ are considered;
we study only positive values of $k$ for two reasons.
Mathematically there is a symmetry between positive and negative momenta
such that negating $k$ conjugates the reflection and transmission
coefficients; physically if one considers propagating wavepackets with
tightly peaked momenta instead of un-normalizable plane waves, the
negative momenta result in the initial state traveling away from the
scattering graph instead of toward it, under the definitions we use.
Note that in Ref.~\cite{Childs:2009b} the graphs of Fig.~\ref{fig:Childsgraphs}
are utilized at momentum $k=-\pi/4$ because the Hamiltonian employed therein
is equal to the adjacency matrix of the graph, rather than its negative.
We find that the number of graphs meeting the requirements for implementing
single-qubit rotations does indeed grow exponentially with $n$.
The most prolific momentum we investigate is $k=2\pi/3$, which results in a
total of 98 unique unitary operations on graphs of nine or fewer vertices.

There is a supplemental data file associated with this manuscript, available
on the arXiv pre-print server
\cite{SupMat}, and our main focus here is the analysis of those
data.
This manuscript is organized as follows. The mathematical formalism underlying
the scattering of plane waves by graphs is briefly reviewed in 
Sec.~\ref{sec:formalism}, followed by a description of the numerical strategy
to enumerate graphs. The results of the survey are given in 
Sec.~\ref{sec:results}, and are summarized in Sec.~\ref{sec:conclusions}.

\section{Formalism}
\label{sec:formalism}

\subsection{Graph Scattering}

The theory underpinning the scattering of plane waves by graphs has been 
discussed in detail by Varbanov and Brun~\cite{Varbanov:2009}, so it will only
be sketched here. One considers a finite graph $G=\{E,V\}$ with $V$ a set 
of $n$ vertices and $E\subseteq V\times V$ a set of edges defined by the graph
adjacency matrix $A^G$.
Each vertex $v\in V$ has attached to it $M_v$ semi-infinite tails ($M_v$ may
vanish for as many as $n-1$ of the vertices of $G$),
with the adjacency matrix of tail $m$ on vertex $v$ defined as
$\left[A^T_{vm}\right]_{ij}=\delta_{i,j+1}
+\delta_{i+1,j}$ where $i,j\in\{1,2,3,\ldots\}$.
Vertices on the tails are 
denoted $\ket{j_{vm}}$, with the attachment vertex on the graph labeled 
$\ket{0_{vm}}$ where the index $m$ is kept only to match the tail labeling,
i.e.\ $\ket{0_{vm}}\equiv\ket{v}$.
The adjacency matrix associated with the connection of the tails to the graph
is $A^C_{vm}=\ketbra{0_{vm}}{1_{vm}}+\ketbra{1_{vm}}{0_{vm}}$, and the
total adjacency matrix for the scattering system as a whole is
\begin{equation}
  A = A^G + \sum_{v\in V}\sum_{m=1}^{M_v}\left(A^T_{vm}+A^C_{vm}\right).
\end{equation}
To describe single-qubit unitary transformations we are interested
only in those cases with four tails --- two input and two output ---
attached to $G$. That is, we require
\begin{equation}\label{eq:MvSumTo4}
  \sum_{v\in G}M_v=4.
\end{equation}

For the moment though, suppose that the graph $G$ consists of a single point.
In this
case, two semi-infinite tails corresponding to the input and output states
fuse into a single infinite line. The (unnormalized) eigenstates of the 
graph Hamiltonian $H\equiv-A$ are plane waves 
\begin{equation}
  \ket{k}=\ket{0}+\sum_{j=1}^{\infty}
    \left(e^{-\imath kj}\ket{j_{00}}+e^{\imath kj}\ket{j_{01}}\right)
\end{equation}
with eigenvalue $\epsilon_k=-2\cos(k)$, where $k\in\left[-\pi,\pi\right)$ is 
the momentum quantum number. One can envisage the input
as a wavepacket strongly peaked at momentum $k$ scattering off a 
single point, resulting in an identical output wavepacket. 

Now consider an arbitrary
number of tails connected to an arbitrary finite graph $G$, again defined by
adjacency matrix $A$ and corresponding Hamiltonian $H=-A$.One again assumes that the input and output states are plane-wave eigenstates 
of their respective semi-infinite lines, with eigenvalue $2\cos(k)$; that is, 
the scattering has not changed the value of $k$,
equivalent to the conservation 
of momentum and energy. In principle the graph could mix momenta on different 
input and output tails; this possibility is not included in the analysis, 
because it would imply that the quantum information encoded in the 
wavepackets would travel down the tail at different rates.

Consider a momentum eigenstate $\ket{k,vm}$ with incoming component only on
tail $m$ of vertex $v$, and possible outgoing components on every attached tail
including the incoming one.
This leads to transmission and reflection coefficients,
$t_{v'm',vm}$ and $r_{vm}$ respectively, defined by the 
expressions
\begin{eqnarray}
	\left\langle j_{vm}|k,{vm}\right\rangle
		&=&
		e^{-\imath kj}+r_{vm}(k)e^{\imath kj}, \label{eq:DefineR} \\
	\left\langle j_{v'm'}|k,{vm}\right\rangle
		&=&
		t_{v'm',vm}(k)e^{\imath kj}. \label{eq:DefineT}
\end{eqnarray}
The first and second terms in Eq.~(\ref{eq:DefineR}) represent the incoming 
and reflected component of the state respectively. The term on the right hand 
side of Eq.~(\ref{eq:DefineT}) represents the component of the incoming state 
that is transmitted to the other tails (for which $v'\neq v$ or $m'\neq m$ ). 
These are Eqs.~(4) and (5) of Varbanov and Brun~\cite{Varbanov:2009}. 

The momentum eigenstates $\ket{k,{vm}}$ can be expressed in terms of their
components on the graph vertices $\ket{k,{vm}}^G$ and the tails:
\begin{eqnarray}
\ket{k,{vm}} &=& \ket{k,{vm}}^{G}+\sum_{j=1}^{\infty}(e^{-\imath kj}
+r_{vm}e^{\imath kj})\ket{j_{vm}}\nonumber \\
&&+ \left.\sum_{v',m'}\right.^{\prime}t_{v'm',vm}\sum_{j=1}^{\infty}
e^{\imath kj}\ket{j_{v'm'}} ,
\label{eq:KVM}
\end{eqnarray}
where the prime on the second sum indicates that all possible values of $v'$ 
and $m'$ for which $v'\neq v$ or $m'\neq m$ are being summed over. 
Enforcing the condition that
${H}\ket{k,{vm}}=-2\cos(k)\ket{k,{vm}}$ and simplifying, one obtains that
$\ket{k,vm}^G$ must satisfy
\begin{eqnarray}
&&\left(A^{G}-2\cos(k)+e^{\imath k}{\sum_{v'}}M_{v'}\left|v'\right\rangle 
\left\langle v'\right|\right)\left|k,{vm}\right\rangle ^{G}\nonumber \\
&&\qquad\qquad\qquad\qquad\qquad\qquad = 2\imath \sin(k)\left|v\right\rangle.
\label{eq:main}
\end{eqnarray}
This is Eq.~(10) of Varbanov and Brun~\cite{Varbanov:2009}. The solution to 
the scattering problem consists of determining the vector $\ket{k,{vm}}^G$ and 
from it obtaining the reflection and transmission coefficients. Because the 
reflection and transmission coefficients at the attachment points must be the
same as on the tails,  Eqs.~(\ref{eq:DefineR}) and (\ref{eq:DefineT}) can be
expressed more conveniently in terms of $\ket{k,vm}^G$ and the graph vertex
states as
\begin{eqnarray}
\left\langle v|k,{vm}\right\rangle^G&=&1+r_{vm}, \label{eq:reflect}\\
\left\langle v'|k,{vm}\right\rangle^G&=&t_{v'm',vm}. \label{eq:transmit}
\end{eqnarray}

In order to interpret the scattering event as implementing a single-qubit
operation, one assumes that the finite graph $G$ has four semi-infinite tails
attached. Two of the tails are considered as inputs and two as outputs.
A continuous-time quantum walker in a wavepacket state with momentum
tightly peaked about $k$ propagates toward $G$ along the two input tails.
It scatters from an initial superposition supported on these tails to a
superposition with support on the two output tails. Any relative phase shift
or change in probability amplitude between the two tails corresponds to
a unitary transformation of the input computational state.
In general, a momentum eigenstate incoming on the first input, $vm=0_\In$,
can be thought of as mapping incoming to outgoing states according to
\begin{subequations}\label{eq:generalScattering}
\begin{align}
  \ket{k,0_\In}
      \mapsto&\ 
      r_{0_\In}(k)\ket{k,0_\In} + t_{1_\In,0_\In}(k)\ket{k,1_\In} \nonumber \\
      &\quad+\sum_{i=0}^1 t_{i_\Out,0_\In}(k)\ket{k,i_\Out}.
\end{align}
Similarly, a momentum state with incoming component on the second input,
$vm=1_\In$, scatters to outgoing components as
\begin{align}
  \ket{k,1_\In}
      \mapsto&\ 
    r_{1_\In}(k)\ket{k,1_\In} + t_{0_\In,1_\In}(k)\ket{k,0_\In} \nonumber \\
      &\quad+\sum_{i=0}^1 t_{i_\Out,1_\In}(k)\ket{k,i_\Out},
\end{align}
\end{subequations}
so in order for these scattering processes to correspond to a unitary operation
mapping $\ket{k,0_\In}$ and $\ket{k,1_\In}$ to superpositions of
$\ket{k,0_\Out}$ and $\ket{k,1_\Out}$ we must identify combinations of graphs
$G$; tail attachment points $\ket{0_\In}$, $\ket{1_\In}$, $\ket{0_\Out}$, and
$\ket{1_\Out}$; and momenta $k$ such that there is neither reflection along
either input tail nor transmission from one input tail to the other.
That is, we require
\begin{equation}\label{eq:unitaryCond}
  r_{0_\In} = r_{1_\In} = t_{0_\In,1_\In} = t_{1_\In,0_\In} = 0.
\end{equation}
When the conditions Eq.~\eqref{eq:unitaryCond} are met, the transformations
Eq.~\eqref{eq:generalScattering} simplify to
\begin{subequations}
\begin{align}
  \ket{k,0_\In}
      &\mapsto
      t_{0_\Out,0_\In}(k)\ket{k,0_\Out}
      + t_{1_\Out,0_\In}(k)\ket{k,1_\Out}, \\
  \ket{k,1_\In}
      &\mapsto
      t_{0_\Out,1_\In}(k)\ket{k,0_\Out}
      + t_{1_\Out,1_\In}(k)\ket{k,1_\Out},
\end{align}
\end{subequations}
i.e.\ an arbitrary input vector $\alpha\ket{k,0_\In}+\beta\ket{k,1_\In}$ is
transformed by the operator:
\begin{equation}
\hat{O}=\left(
\begin{matrix}
t_{0_{\Out},0_{\In}} & t_{0_{\Out},1_{\In}}\\
t_{1_{\Out},0_{\In}} & t_{1_{\Out},1_{\In}}
\end{matrix}
\right).
\label{eq:Operation}
\end{equation}
We must now show that this operator is unitary. To simplify notation, let us
label the two input tails by $1$ and $2$, and the output tails by $3$ and $4$.
A general scattering process on a graph with four attached tails yields an
S-matrix
\begin{align}
  S = \sum_{i=1}^4\left(
        r_{i}\ketbra{\tau_i}{\tau_i}
        +\sum_{\substack{j=1\\j\neq i}}^4
          t_{ij}\ketbra{\tau_i}{\tau_j}
      \right),
\end{align}
where $\ket{\tau_i}$ corresponds to the state on tail $i$,
which Varbanov and Brun show is unitary \cite{Varbanov:2009}.
Our restriction that we consider only graphs
satisfying Eq.~\eqref{eq:unitaryCond} does not change the unitarity of $S$
in those cases, and corresponds to $r_{1}=r_{2}=t_{12}=t_{21}=0$.
We then assume an initial state with support only on the input tails,
$\ket{\psi_0}=\alpha\ket{\tau_1}+\beta\ket{\tau_2}$,
with $|\alpha|^2+|\beta|^2=1$.
The restricted S-matrix acts on such a state as
\begin{equation}
  S\ket{\psi_0}
  =
  \left(\alpha t_{31} + \beta t_{32}\right)\ket{\tau_3}
  +\left(\alpha t_{41} + \beta t_{42}\right)\ket{\tau_4},
\end{equation}
but this is precisely the action of $\hat{O}$ on 
$\alpha\ket{k,0_\In}+\beta\ket{k,1_\In}$
under the identification
$\ket{\tau_i}\leftrightarrow\ket{k,i}$ with
$1\leftrightarrow 0_\In$,
$2\leftrightarrow 1_\In$, $3\leftrightarrow 0_\Out$, and
$4\leftrightarrow 1_\Out$.
Since $S$ is unitary in general, it remains so when restricted to
the cases of interest that define $\hat{O}$.

\subsection{Enumeration of Graphs}
The number of non-isomorphic simple graphs on $n$ vertices grows
super-exponentially 
with $n$, and for each graph the number of ways to attach
four tails goes as $n^4$.
Using the \texttt{geng} (\emph{gen}erate \emph{g}raphs) software package,
part of the popular \texttt{nauty}
(\emph{n}o \emph{aut}omorphisms, \emph{y}es?) suite of graph theoretic
tools \cite{McKay},
we first generate the adjacency matrices $A^G$ of all non-isomorphic simple
graphs on $n$ vertices for $n\in\{1,2,3,\ldots,9\}$.  
For each graph we then enumerate the set of all combinations of
attachment points for four tails, each of which defines a set
$\{M_1,M_2,\ldots,M_n\}$ satisfying Eq.~\eqref{eq:MvSumTo4}.

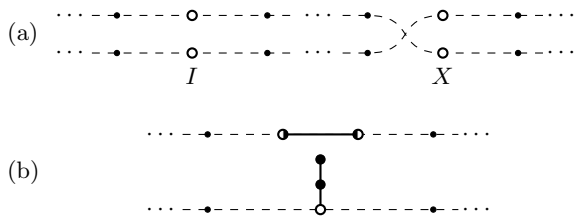
\begin{figure}
  \begin{center}
    \subfloat{\label{subfig:trivialIandX}
        \begin{tikzpicture}
        \draw[tail] (2,0.25) -- (2.333,0.25)
                      .. controls (3,0.25) and (2.667,-0.25)
                      .. (3.333,-0.25) -- (4.667,-0.25);
        \draw[tail] (2,-0.25) -- (2.333,-0.25)
                      .. controls (3,-0.25) and (2.667,0.25)
                      .. (3.333,0.25) -- (4.667,0.25);
        \foreach \y in {0.25,-0.25} {
          \node at (-1.6,\y) {$\cdots$};
          \draw[tail] (-1.333,\y) -- (1.333,\y);
          \node[tailv] at (-1,\y) {};
          \node[tailv] at (1,\y) {};
          \node[vtx] at (0,\y) {};
          \vtxIO{0,\y}
          \node at (1.7,\y) {$\cdots$};
          \node[tailv] at (2.333,\y) {};
          \node[tailv] at (4.333,\y) {};
          \node[tailv] at (3.333,\y) {};
          \vtxIO{3.333,\y}
          \node at (4.934,\y) {$\cdots$};
        };
        \node at (-0.005,-0.55) {$I$};
        \node at (3.328,-0.55) {$X$};
        \node at (-2.25,0) {\subref{subfig:trivialIandX}};
      \end{tikzpicture}
    }\\
    \subfloat{\label{subfig:nonTrivialOn5}
      \begin{tikzpicture}
        \foreach \y in {0.5,-0.5} {
          \node[tailv] at (-1.5,\y) {};
          \node[tailv] at (1.5,\y) {};
          \draw[tail] (-1.833,\y) -- (1.833,\y);
          \node at (-2.1,\y) {$\cdots$};
          \node at (2.1,\y) {$\cdots$};
        };
        \node[vtx] (v0) at (-0.5,0.5) {};
        \node[vtx] (v1) at (0.5,0.5) {};
        \node[vtx] (v2) at (0,-0.5) {};
        \node[vtx] (v3) at (0,-0.167) {};
        \node[vtx] (v4) at (0,0.167) {};
        \vtxIn{v0}
        \vtxOut{v1}
        \vtxIO{v2}
        \vtx{v3}
        \vtx{v4}
        \draw[edge] (v0) -- (v1);
        \draw[edge] (v2) -- (v4);
        \node at (-3.95,0) {\subref{subfig:nonTrivialOn5}};
        \node at (3.234,0) {\hphantom{$\cdots$}};
      \end{tikzpicture}
    }
  \end{center}
  \caption{\label{fig:trivialGraphs}
    \subref{subfig:trivialIandX} Given two isolated vertices off of which to
    scatter, the identity and $\textsc{swap}$ gates, $I$ and $X$, are trivial
    to construct at any momentum
    (supplemental material IDs 180 and 179, respectively, at $k=\pi/4$). 
    \subref{subfig:nonTrivialOn5} With five vertices it becomes possible to
    implement gates other than $I$ and $X$. This five-vertex graph implements
    $R_Z(-\pi/2)$ and has an effective length of $1$.
    (Supplemental material ID 1458, $k=\pi/2$.)
  }
\end{figure}

One can see by symmetry arguments that the one-vertex graph with four
tails attached cannot transmit from one of the tails to only two of
the remaining three, so at least two vertices are required to implement
a computational unitary gate.
With two or more vertices available, it is trivial to implement the
identity gate and the $\textsc{swap}$ or $X$ gate
by constructing a graph with two isolated vertices and
attaching an input and an output tail to each one, as exemplified in
Fig.~\ref{subfig:trivialIandX}. If the computational labels (i.e.\ $\ket{0}$
and $\ket{1}$) agree on each vertex then an identity is performed;
otherwise the $X$ gate is applied.
The first unitaries that cannot be trivially produced in this manner appear
when there are five vertices. An example of such a graph can be found in
Fig.~\ref{subfig:nonTrivialOn5}.

For each momentum value $k$ under consideration, for each adjacency
matrix $A^G$ and attachment set $\{M_v\}$,
Eq.~\eqref{eq:main} produces a set of $n$
linear equations to be solved for the coefficients of $\ket{k,vm}^G$.
These in turn yield the reflection and transmission coefficients from
Eqs.~\eqref{eq:reflect} and \eqref{eq:transmit}.
Due to the large number of such combinations of $k$, $\{M_v\}$, and $A^G$
(of order $10^9$ under the chosen constraints), we solve these equations
numerically using the GNU Scientific Library \cite{GSL}
and LAPACK \cite{LAPACK} software libraries.
If a particular combination results in reflection and transmission
coefficients satisfying Eq.~\eqref{eq:unitaryCond} then there is zero
transmission from the input tail to one of the three possible output
tails. In this case, we solve the equations again with the incoming
portion of the momentum eigenstate instead on this unused tail.
If the conditions \eqref{eq:unitaryCond} are again satisfied, with the
same two tails supporting the output, then the current combination
implements a unitary transformation $U$ defined by the operator
\eqref{eq:Operation}.

The final step is to determine the effective length through the graph
from each input to each output. If paths of different length exist, then
a quantum walker initially in a superposition of one wavepacket on either
input tail would acquire a spatial shift between the two output wavepackets
after traversing the graph. The effective length to output
$j$ from input $i$ is defined as \cite{Childs:2009b}
\begin{equation}\label{eq:effLen}
  \ell_{ji}(k)
  =
  \frac{d}{dk}\arg t_{j_{\Out},i_{\In}}(k).
\end{equation}
This can be understood when one considers the case of a single input
attached to one end of a linear graph of $L$ segments, with a single output
tail at the other end. Clearly the transmission from one end to the other
has unit magnitude at all momenta, and the phase difference between the ends
results in a transmission coefficient of
$t(k)=e^{\imath k L}$. Equation \eqref{eq:effLen} then yields
$\ell(k)=L$, as expected.
We calculate this derivative numerically on each candidate graph, using
a nine-point stencil, and require that the four values $\ell_{ji}(k)$ agree
whenever $t_{j_\Out,i_\In}\neq 0$.

Our goal is to determine any potential computational advantage to
increasing the number of vertices in $G$, so we record the current system
only if $U$ meets all of the stated criteria and has not been found at the
same momentum and effective length on a smaller graph.
If it has been previously found, we keep track of how many times it has
appeared.

\section{Results}\label{sec:results}

\begin{figure}[t]
  \begin{center}
    \subfloat{
      \begin{tikzpicture}
        \node[inner sep=0pt,above right]
          {\includegraphics[width=222pt]{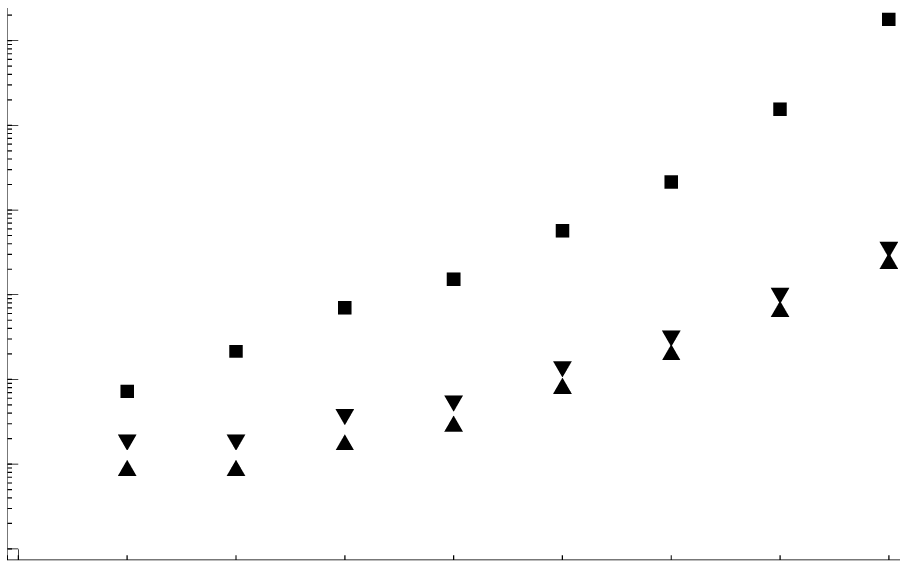}};
        \foreach \y in {1,...,9} {
          \node at (0.94*\y-0.76,0.2) {\y};
        };
        \node[left] at (0.1,0*0.73+0.6) {1\hphantom{$^2$}};
        \node[left] at (0.1,1*0.73+0.6) {10\hphantom{$^2$}};
        \foreach \y in {2,...,6} {
          \node[left] at (0.1,0.73*\y+0.6) {$10^{\y}$};
        %
        %
        \draw[line width=0.05pt] (0.35,5.2)
          --++(4.1,0) --++(0,-1.05) --++(-4.1,0) -- cycle;
        };
        \lgndSquareFilled{0.5}{5}{0.12}
        \node[right] at (0.5,5)
          {\scriptsize Graphs resulting in unitaries};
        \lgndTriDownFilled{0.5}{4.667}{0.12}
        \node[right] at (0.5,4.667)
          {\scriptsize Distinct unitaries};
        \lgndTriUpFilled{0.5}{4.333}{0.12}
        \node[right] at (0.5,4.333)
          {\scriptsize Distinct non-identity unitaries};
      \end{tikzpicture}
    }\\
    \subfloat{
      \begin{tikzpicture}
        \node[inner sep=0pt,above right]
          {\includegraphics[width=222pt]{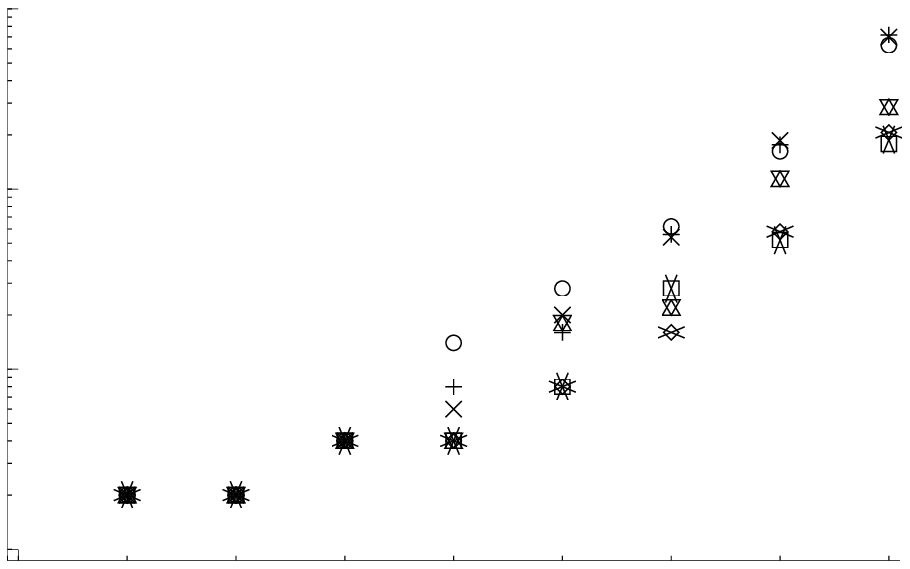}};
        \foreach \y in {1,...,9} {
          \node at (0.94*\y-0.76,0.2) {\y};
        };
        \node[left] at (0.1,0*1.55+0.55) {1\hphantom{$^2$}};
        \node[left] at (0.1,1*1.55+0.55) {10\hphantom{$^2$}};
        \foreach \y in {2,3} {
          \node[left] at (0.1,1.55*\y+0.55) {$10^{\y}$};
        };
        %
        %
        \draw[line width=0.05pt] (0.35,5.15)
          --++(1.9,0) --++(0,-1.31) --++(-1.9,0) -- cycle;
        %
        \lgndTriDown{0.5}{5}{0.12}
        \node[right] at (0.5,5) {$\scriptstyle\pi/4$};
        \lgndPlus{0.5}{4.75}{0.12}
        \node[right] at (0.5,4.75) {$\scriptstyle\pi/3$};
        \lgndCircle{0.5}{4.5}{0.12}
        \node[right] at (0.5,4.5) {$\scriptstyle\pi/2$};
        \lgndTimes{0.5}{4.25}{0.12}
        \node[right] at (0.5,4.25) {$\scriptstyle2\pi/3$};
        \lgndTriUp{0.5}{4}{0.12}
        \node[right] at (0.5,4) {$\scriptstyle3\pi/4$};
        %
        %
        \lgndSquare{1.5}{5}{0.12}
        \node[right] at (1.5,5) {$\scriptstyle\pi/5$};
        \lgndBowtie{1.5}{4.75}{0.12}
        \node[right] at (1.5,4.75) {$\scriptstyle2\pi/5$};
        \lgndTeepee{1.5}{4.5}{0.12}
        \node[right] at (1.5,4.5) {$\scriptstyle3\pi/5$};
        \lgndDiamond{1.5}{4.25}{0.12}
        \node[right] at (1.5,4.25) {$\scriptstyle4\pi/5$};
        \node at (3.94,-0.2) {Vertices};
      \end{tikzpicture}
    }
  \end{center}
  \caption{\label{fig:CountVsN}
    (Top)
    Total number of graphs resulting in unitary operators, as well as the
    number of distinct unitaries found as a function of number of 
    vertices. Here operations are considered distinct if they arise from
    graphs with different effective lengths or at different momenta.
    (Bottom)
    Number of distinct unitaries, as presented in the upper plot,
    separated by momentum.
  }
\end{figure}

The number of non-isomorphic simple graphs on $n$ vertices, $N_n$, does
not have a simple closed-form solution. Nevertheless, the counts are well
documented for many values of $n$ (see, for example,
Ref.~\cite{Sloane:A000088} and references therein), and
the total number of such graphs on nine or fewer vertices
is $288\,266$.
For each graph on $n$ vertices, the number of distinct ways to attach
$k$ tails is
\begin{equation}
  w_n = \binom{n+k-1}{k},
\end{equation}
and we have chosen to investigate scattering at nine different momenta
with four attached tails.
This results in
\begin{equation}
  9\sum_{n=1}^9 N_n w_n = 1\,262\,489\,148
\end{equation}
combinations of momenta, attachment points, and graphs to be numerically 
examined.  Of these, it turns out that $1\,960\,316$ have the properties 
required to implement a computational gate. 
Figure \ref{fig:CountVsN} shows that the number of unitary and non-identity
operations both increase super-exponentially in the number of vertices, 
following the growth of the total number of graphs, and that this is true at 
each momentum investigated.

There is significant redundancy 
within these gates however, where many different graphs yield the same unitary 
for a given value of the momentum and effective length. This reduces the total 
number of distinct unitary operations to $3380$. The adjacency matrices for 
these graphs, the relevant momentum and 
attachment points, and the resulting unitary gates arising, are given 
explicitly in the supplemental online information accompanying this
work \cite{SupMat}. Of the graphs producing
these, 2496 yield gates other than identities; that said, it is 
important to keep in mind that an identity counterpart is required for each 
non-identity gate in order to ensure universality so having a sufficiently
large number of identity gates is essential.

\begin{figure}
  \begin{center}
    \begin{tikzpicture}
      \node[inner sep=0pt,above right]
        {\includegraphics[width=228pt]{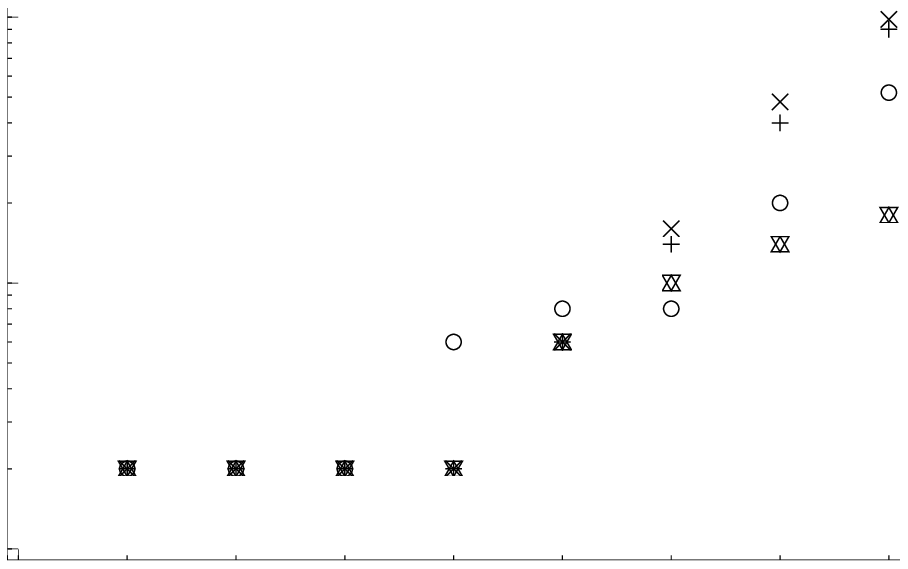}};
      \foreach \y in {1,...,9} {
        \node at (0.965*\y-0.783,0.25) {\y};
      };
      \node[left] at (0.15,0*2.36+0.6) {1};
      \node[left] at (0.15,1*2.36+0.6) {10};
      \node[left] at (0.15,2*2.36+0.6) {100};
      \draw[line width=0.05pt] (0.35,5.15)
          --++(1.9,0) --++(0,-0.786) --++(-1.9,0) -- cycle;
        %
        \lgndTriDown{0.5}{5}{0.12}
        \node[right] at (0.5,5) {$\scriptstyle\pi/4$};
        \lgndPlus{0.5}{4.75}{0.12}
        \node[right] at (0.5,4.75) {$\scriptstyle\pi/3$};
        \lgndCircle{0.5}{4.5}{0.12}
        \node[right] at (0.5,4.5) {$\scriptstyle\pi/2$};
        %
        \lgndTimes{1.5}{5}{0.12}
        \node[right] at (1.5,5) {$\scriptstyle2\pi/3$};
        \lgndTriUp{1.5}{4.75}{0.12}
        \node[right] at (1.5,4.75) {$\scriptstyle3\pi/4$};
        \node at (3.94,-0.15) {Vertices};
    \end{tikzpicture}
  \end{center}
  \caption{\label{fig:UwithN}
    Number of available unitaries provided by
    increasing the number of vertices in the scattering graph, for
    the five $k$ values at which an increase was found.
    Here we do not consider unitaries to be distinct if they arise
    at different effective lengths but otherwise perform the same
    transformation.
  }
\end{figure}

Only those graphs which result in unitary operations at the same momentum
value can be combined under the current model to effect a quantum computation,
since the momentum of the quantum walker is fixed.
The effective lengths of the constituent graphs are unimportant, as long as
each graph can be paired with an identity operation of the same length.
There are $262$ graphs which do not have commensurate identity graphs,
leaving $3118$ candidates for inclusion in computational sets.
There is further redundancy among these graphs,
since two graphs performing the same operation with different effective lengths
at the same momentum
are for our purposes equivalent if each has a corresponding identity operation.
Taking this into account reduces the number of potentially useful operations to
$284$. The number of resulting unitaries is shown in Fig.~\ref{fig:UwithN} as a 
function of the number of vertices for the momentum values investigated.
There are 16 unitaries that can be produced by graphs on five vertices,
eight of which are new in that they cannot be produced by four or fewer
vertices. Similarly, there
are 24 new unitaries on six vertices, and 30 more on seven. These 62 unitary
operators are produced by only 15 graphs, combined with various tail attachment
and momentum configurations. One of these graphs has two variations of tail
attachment points that lead to non-isomorphic infinite graphs once the tails
are included. These 16 graphs are drawn explicitly in
Fig.~\ref{fig:Gs5-7v}.
\tikzset{x=1.5cm,y=1.5cm}
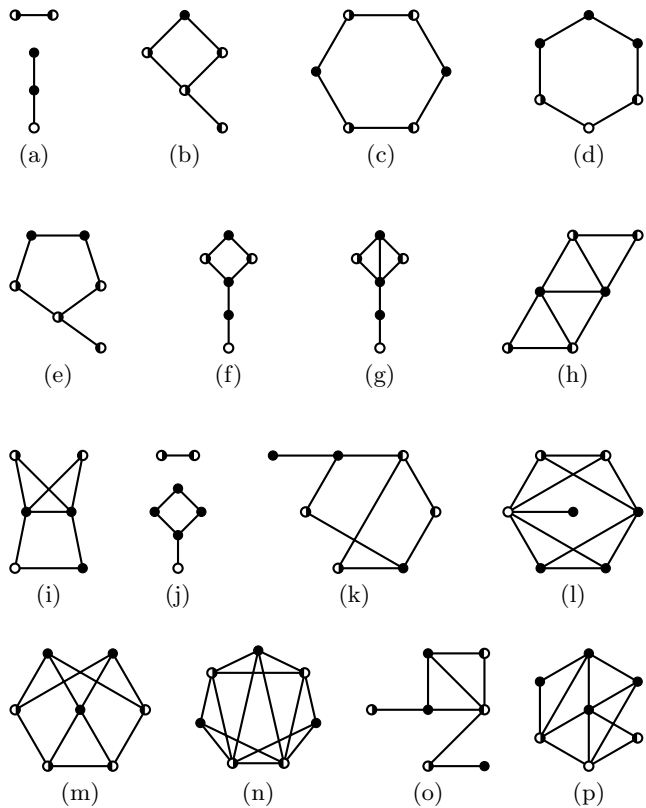
\begin{figure}
  \begin{center}
    %
    %
    \subfloat{\label{subfig:v5-1}
      \begin{tikzpicture}
        \draw[edge] (-0.167,1) -- (0.167,1);
        \draw[edge] (0,0) -- (0,0.667);
        \vtxIn{-0.167,1}
        \vtxOut{0.167,1}
        \vtxIO{0,0}
        \vtx{0,0.333}
        \vtx{0,0.667}
        \node at (0,-0.25) {\subref{subfig:v5-1}};
      \end{tikzpicture}
    }\hfill
    \subfloat{\label{subfig:v5-2}
      \begin{tikzpicture}
        \draw[edge] (0.333,0.333) -- (0,0.667) -- (0.333,1)
          -- (0.667,0.667) -- (0.333,0.333) -- (0.667,0);
        \vtxIn{0,0.667}
        \vtxIn{0.333,0.333}
        \vtxOut{0.667,0.667}
        \vtxOut{0.667,0}
        \vtx{0.333,1}
        \node at (0.333,-0.25) {\subref{subfig:v5-2}};
      \end{tikzpicture}
    }\hfill
    %
    %
    \subfloat{\label{subfig:v6-1}
      \begin{tikzpicture}
        \draw[edge] (0:0.577) \foreach \x in {60,120,...,359} {
          -- (\x:0.577)
        } -- cycle;
        \vtxIn{240:0.577}
        \vtxIn{120:0.577}
        \vtxOut{-60:0.577}
        \vtxOut{60:0.577}
        \foreach \x in {0,180} {
          \vtx{\x:0.577}
        }
        \node at (0,-0.75) {\subref{subfig:v6-1}};
      \end{tikzpicture}
    }\hfill
    \subfloat{\label{subfig:v6-2}
      \begin{tikzpicture}
        \draw[edge] (-90:0.5) \foreach \x in {60,120,...,359} {
          -- (-90+\x:0.5)
        } -- cycle;
        \vtxIn{210:0.5}
        \vtxOut{-30:0.5}
        \vtxIO{-90:0.5}
        \foreach \x in {30,90,150} {
          \vtx{\x:0.5}
        }
        \node at (0,-0.75) {\subref{subfig:v6-2}};
      \end{tikzpicture}
    }\\
    \vspace{1em}
    \subfloat{\label{subfig:v6-3}
      \begin{tikzpicture}
        \draw[edge] (-90:0.4) \foreach \x in {72,144,...,359} {
          -- (-90+\x:0.4)
        } -- cycle;
        \draw[edge] (0,-0.4) -- (0.380,-0.673);
        \vtxIn{0,-0.4}
        \vtxIn{198:0.4}
        \vtxOut{-18:0.4}
        \vtxOut{0.380,-0.673}
        \vtx{54:0.4}
        \vtx{126:0.4}
        \node at (0,-0.673-0.25) {\subref{subfig:v6-3}};
      \end{tikzpicture}
    }\hfill
    \subfloat{\label{subfig:v6-4}
      \begin{tikzpicture}
        \draw[edge] (0,0) -- (0,0.586) -- (-0.207,0.793)
          -- (0,1) -- (0.207,0.793) -- (0,0.586);
        \vtxIO{0,0}
        \vtxIn{-0.207,0.793}
        \vtxOut{0.207,0.793}
        \vtx{0,0.293}
        \vtx{0,0.586}
        \vtx{0,1}
        \node at (0,-0.25) {\subref{subfig:v6-4}};
      \end{tikzpicture}
    }\hfill
    \subfloat{\label{subfig:v6-5}
      \begin{tikzpicture}
        \draw[edge] (0,0) -- (0,0.586) -- (-0.207,0.793)
          -- (0,1) -- (0.207,0.793) -- (0,0.586) -- (0,1);
        \vtxIO{0,0}
        \vtxIn{-0.207,0.793}
        \vtxOut{0.207,0.793}
        \vtx{0,0.293}
        \vtx{0,0.586}
        \vtx{0,1}
        \node at (0,-0.25) {\subref{subfig:v6-5}};
      \end{tikzpicture}
    }\hfill
    \subfloat{\label{subfig:v6-6}
      \begin{tikzpicture}
        \draw[edge] (0.577,0) -- (0,0) -- (0.577,1) -- (1.155,1)
          -- (0.577,0) -- (0.287,0.5) -- (0.866,0.5) -- (0.577,1);
        \vtxIn{0,0}
        \vtxIn{0.577,1}
        \vtxOut{0.577,0}
        \vtxOut{1.155,1}
        \vtx{0.287,0.5}
        \vtx{0.866,0.5}
        \node at (0.577,-0.25) {\subref{subfig:v6-6}};
      \end{tikzpicture}
    }\\
    \vspace{1em}
    \subfloat{\label{subfig:v6-7}
      \begin{tikzpicture}
        \draw[edge] (0.1,0.5) -- (0,0) -- (0.6,0) -- (0.5,0.5)
          -- (0.1,0.5) -- (0,1) -- (0.5,0.5) -- (0.6,1) -- (0.1,0.5);
        \vtxIn{0,1}
        \vtxOut{0.6,1}
        \vtxIO{0,0}
        \vtx{0.1,0.5}
        \vtx{0.5,0.5}
        \vtx{0.6,0}
        \node at (0.3,-0.25) {\subref{subfig:v6-7}};
      \end{tikzpicture}
    }\hfill
    %
    %
    \subfloat{\label{subfig:v7-1}
      \begin{tikzpicture}
        \draw[edge] (0,0) -- (0,0.293) -- (-0.207,0.5) -- (0,0.707)
           -- (0.207,0.5) -- (0,0.293);
        \draw[edge] (-0.146,1) -- (0.146,1);
        \vtxIn{-0.146,1}
        \vtxOut{0.146,1}
        \vtxIO{0,0}
        \vtx{0,0.293}
        \vtx{0,0.707}
        \vtx{-0.207,0.5}
        \vtx{0.207,0.5}
        \node at (0,-0.25) {\subref{subfig:v7-1}};
      \end{tikzpicture}
    }\hfill
    \subfloat{\label{subfig:v7-2}
      \begin{tikzpicture}
        \draw[edge] (60:0.577) -- (240:0.577)
          \foreach \x in {60,120,...,359} {
          -- (-120+\x:0.577)
        } -- (-60:0.577);
        \draw[edge] (120:0.577) --++(-0.577,0);
        \vtxIn{180:0.577}
        \vtxIn{240:0.577}
        \vtxOut{0:0.577}
        \vtxOut{60:0.577}
        \foreach \x in {-60,120} {
          \vtx{\x:0.577}
        }
        \vtx{-0.866,0.5}
        \node at (-0.145,-0.75) {\subref{subfig:v7-2}};
      \end{tikzpicture}
    }\hfill
    \subfloat{\label{subfig:v7-3}
      \begin{tikzpicture}
        \draw[edge] (0:0.577)
          \foreach \x in {60,120,...,359} {
            -- (\x:0.577)
        } -- cycle;
        \draw[edge] (120:0.577) -- (0:0.577) -- (-120:0.577);
        \draw[edge] (60:0.577) -- (180:0.577) -- (-60:0.577);
        \draw[edge] (180:0.577) -- (0,0);
        \vtxIO{180:0.577}
        \vtxIn{120:0.577}
        \vtxOut{60:0.577}
        \foreach \x in {0,-60,-120} {
          \vtx{\x:0.577}
        }
        \vtx{0,0}
        \node at (0,-0.75) {\subref{subfig:v7-3}};
      \end{tikzpicture}
    }\\
    \subfloat{\label{subfig:v7-4}
      \begin{tikzpicture}
        \draw[edge] (60:0.577) \foreach \x in {360,300,...,120} {
          -- (\x:0.577)
        };
        \draw[edge] (180:0.577) -- (60:0.577) -- (0,0)
          -- (120:0.577) -- (0:0.577);
        \draw[edge] (240:0.577) -- (0,0) -- (300:0.577);
        \vtxIn{180:0.577}
        \vtxIn{240:0.577}
        \vtxOut{0:0.577}
        \vtxOut{300:0.577}
        \vtx{0,0}
        \vtx{60:0.577}
        \vtx{120:0.577}
        \node at (0,-0.75) {\subref{subfig:v7-4}};
      \end{tikzpicture}
    }\hfill
    \subfloat{\label{subfig:v7-5}
      \begin{tikzpicture}
        \draw[edge] (90:0.526) \foreach \x in {1,...,7} {
          -- (90+\x*51.429:0.526)
        };
        \draw[edge] (90+2*51.429:0.526) \foreach \x in {4,6,1,3,5} {
          -- (90+\x*51.429:0.526)
        };
        \draw[edge] (90+3*51.429:0.526) -- (90:0.526) -- (90+4*51.429:0.526);
        \foreach \x in {1,3} {
          \vtxIn{90+\x*51.429:0.526}
        }
        \foreach \x in {4,6} {
          \vtxOut{90+\x*51.429:0.526}
        }
        \foreach \x in {0,2,5} {
          \vtx{90+\x*51.429:0.526}
        }
        \node at (0,-0.75) {\subref{subfig:v7-5}};
      \end{tikzpicture}
    }\hfill
    \subfloat{\label{subfig:v7-6}
      \begin{tikzpicture}
        \draw[edge] (1,0) -- (0.5,0) -- (1,0.5) -- (1,1) -- (0.5,1)
          -- (0.5,0.5) -- (1,0.5) -- (0.5,1);
        \draw[edge] (0,0.5) -- (0.5,0.5);
        \vtxIn{0,0.5}
        \vtxIn{0.5,0}
        \vtxOut{1,1}
        \vtxOut{1,0.5}
        \vtx{1,0}
        \vtx{0.5,0.5}
        \vtx{0.5,1}
        \node at (0.5,-0.25) {\subref{subfig:v7-6}};
      \end{tikzpicture}
    }\hfill
    \subfloat{\label{subfig:v7-7}
      \begin{tikzpicture}
        \draw[edge] (30:0.5) \foreach \x in {90,150,...,330} {
          -- (\x:0.5)
        } -- (0,0) -- (30:0.5) -- (270:0.5) -- (90:0.5)
          -- (210:0.5) -- (0,0);
        \vtxIn{210:0.5}
        \vtxOut{-30:0.5}
        \vtxIO{-90:0.5}
        \foreach \x in {30,90,150} {
          \vtx{\x:0.5}
        }
        \vtx{0,0}
        \node at (0,-0.75) {\subref{subfig:v7-7}};
      \end{tikzpicture}
    }
  \end{center}
  \caption{\label{fig:Gs5-7v}
    The graphs on five, six, or seven vertices that implement
    single-qubit unitaries which are unavailable on fewer vertices.
    The five-vertex graphs \subref{subfig:v5-1} and \subref{subfig:v5-2}
    result in eight distinct unitaries under different tail attachment
    configurations and different momentum values.
    The six-vertex graphs \subref{subfig:v6-1}-\subref{subfig:v6-7}
    yield a total of 24 unitaries, and the seven-vertex graphs
    \subref{subfig:v7-1}-\subref{subfig:v7-7} lead to 30.
    Note that while \subref{subfig:v6-1} and \subref{subfig:v6-2}
    are isomorphic, they represent two distinct, non-isomorphic
    configurations once tails are attached.    
  }
\end{figure}
\tikzset{x=1cm,y=1cm}

\subsection{Unitary operations by momentum value}
There are four momentum values, $p\pi/5$ for $p=1,2,3,4$, at which the 
only available operations are the identity and Pauli-$X$ operators;
these momenta are clearly not useful for 
our purposes, and are therefore not plotted. It is clear that for each
momentum value that provides a growing number of unitaries, the number of
unique gates continues the exponentially
increasing trend as a function of the number of vertices.

At $k=\pi/4$ we find graphs that implement Pauli-$X$, -$Y$, and -$Z$ gates
as well as the rotations $R_X(\pm\pi/2)$ and $R_Z(p\pi/4)$ for
$p\in\{-3,-2,\ldots,3\}$ (the latter including the identity at $p=0$).
These results contain the graphs identified by Childs \cite{Childs:2009b}
and therefore (re)produce a universal set of gates.
Additionally, we find graphs producing rotations by $-\pi$ about those
axes in the equatorial plane of the Bloch sphere making angles with respect
to the $X$ axis of $p\pi/8$ for $p\in\{1,2,3,5,6,7\}$.
Momentum $k=3\pi/4$ reproduces this same set of 18 gates.
 
\begin{figure}
  \begin{center}
    \begin{tikzpicture}
      \node[inner sep=0pt,above right]
        {\includegraphics[width=228pt]{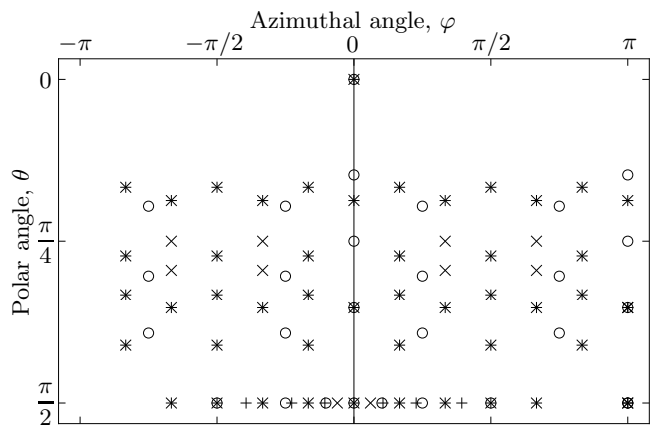}};
      \node at (-0.1,4.92) {$0$};
      \node at (-0.1,2.76) {$\displaystyle\frac{\pi}{4}$};
      \node at (-0.1,0.61) {$\displaystyle\frac{\pi}{2}$};
      \node[rotate=90] at (-0.4,2.76) {Polar angle, $\theta$};
      \node at (0.3,5.4) {$-\pi$};
      \node at (2.143,5.4) {$-\pi/2$};
      \node at (3.985,5.4) {$0$};
      \node at (5.828,5.4) {$\pi/2$};
      \node at (7.67,5.4) {$\pi$};
      \node at (3.985,5.7) {Azimuthal angle, $\varphi$};
    \end{tikzpicture}
  \end{center}
  \caption{\label{fig:axes}
    Distributions of the axes of rotation on the Bloch sphere
    available at momentum values $k=\pi/2$, $\pi/3$, and
    $2\pi/3$.
    That is, there exists at least one non-trivial rotation
    $R_{(\theta,\varphi)}$ about each axis defined by the points
    $(\theta,\varphi)$ on the upper hemisphere of the Bloch sphere
    shown here.
    Symbols are as in Fig.~\ref{fig:UwithN}.
  }
\end{figure}

When the momentum is $k=\pi/2$ we find graphs capable of implementing
$52$ rotations of the Bloch sphere, about $28$ non-parallel axes.
With momenta $k=\pi/3$ and $k=2\pi/3$ we find $90$ and $98$ rotations
about $55$ and $59$ distinct axes, respectively.
The available axes are visualized for these three momentum values in
Fig.~\ref{fig:axes}.
The reflection symmetries about $\phi=0$ and $\phi=\pm\pi/2$ are due
to the different configurations of attachment points available for each
graph.
If a given graph performs a single-qubit rotation about the
$(\theta,\phi)$ axis of the Bloch sphere when the attachment vertices
for $(\ket{0}_\In,\ket{1}_\In,\ket{0}_\Out,\ket{1}_\Out)$ are
$(\ket{1},\ket{2},\ket{3},\ket{4})$, then moving the inputs to the outputs
and vice versa by re-attaching the tails
in the order $(\ket{3},\ket{4},\ket{1},\ket{2})$ results in a rotation
about $(\theta,-\phi)$. Interchanging the inputs with each other and
doing the same for the outputs, i.e.\ re-ordering the tails as
$(\ket{2},\ket{1},\ket{4},\ket{3})$, leads to a rotation about
$(\theta,\pi-\phi)$. This re-ordering is equivalent to conjugating the
original gate by $X$.

\subsection{Rotations by irrational multiples of $\pi$;
            fractional and irrational effective lengths.}
\begin{figure}
  \begin{center}
    \subfloat{\label{subfig:irrationalRotation}
      \begin{tikzpicture} 
        \node at (-4.25,0) {\subref{subfig:irrationalRotation}};
        \node at (3.5,0) {\hphantom{$\cdots$}};
        \node[vtx] (v0) at (0.,-0.75) {};
        \node[vtx] (v1) at (0.4971,-0.5691) {};
        \node[vtx] (v2) at (0.7616,-0.111) {};
        \node[vtx] (v3) at (0.6697,0.41) {};
        \node[vtx] (v5) at (0.2645,0.75) {};
        \node[vtx] (v7) at (-0.2645,0.75) {};
        \node[vtx] (v4) at (-0.6697,0.41) {};
        \node[vtx] (v6) at (-0.7616,-0.111) {};
        \node[vtx] (v8) at (-0.4971,-0.5691) {};
        \vtxIO{v0}
        \vtx{v1}
        \vtx{v2}
        \vtx{v3}
        \vtx{v4}
        \vtxOut{v5}
        \vtx{v6}
        \vtxIn{v7}
        \vtx{v8}
        \draw[edge] (v0) -- (v3);
        \draw[edge] (v0) -- (v4);
        \draw[edge] (v0) -- (v5);
        \draw[edge] (v0) -- (v6);
        \draw[edge] (v0) -- (v7);
        \draw[edge] (v0) -- (v8);
        \draw[edge] (v1) -- (v4);
        \draw[edge] (v1) -- (v5);
        \draw[edge] (v1) -- (v6);
        \draw[edge] (v1) -- (v7);
        \draw[edge] (v1) -- (v8);
        \draw[edge] (v2) -- (v6);
        \draw[edge] (v2) -- (v7);
        \draw[edge] (v2) -- (v8);
        \draw[edge] (v3) -- (v4);
        \draw[edge] (v3) -- (v5);
        \draw[edge] (v3) -- (v7);
        \draw[edge] (v3) -- (v8);
        \draw[edge] (v4) -- (v6);
        \draw[edge] (v4) -- (v7);
        \draw[edge] (v4) -- (v8);
        \draw[edge] (v5) -- (v6);
        \draw[edge] (v5) -- (v8);
        \draw[edge] (v7) -- (v8);
        %
        %
        \node[tailv] (in0) at (-1,0.75) {};
        \node[tailv] at (-2,0.75) {};
        \node[tailv] (in1) at (-1,-0.75) {};
        \node[tailv] at (-2,-0.75) {};
        \node[tailv] (out0) at (1,0.75) {};
        \node[tailv] at (2,0.75) {};
        \node[tailv] (out1) at (1,-0.75) {};
        \node[tailv] at (2,-0.75) {};
        \node at (-2.75,0.75) {$\cdots$};
        \draw[tail] (-2.35,0.75) -- (in0);
        \node at (-2.75,-0.75) {$\cdots$};
        \draw[tail] (-2.35,-0.75) -- (in1);
        \node at (2.75,0.75) {$\cdots$};
        \draw[tail] (2.35,0.75) -- (out0);
        \node at (2.75,-0.75) {$\cdots$};
        \draw[tail] (2.35,-0.75) -- (out1);
        \draw[tail] (in0) -- (v7);
        \draw[tail] (in1) -- (v0);
        \draw[tail] (v5) -- (out0);
        \draw[tail] (v0) -- (out1);
      \end{tikzpicture}
    }\\
    \subfloat{\label{subfig:fractionalLength}
      \begin{tikzpicture}
        \node at (-4.25,0) {\subref{subfig:fractionalLength}};
        \node at (3.5,0) {\hphantom{$\cdots$}};
        \node[vtx] (v0) at (-0.75,0.4) {};
        \node[vtx] (v1) at (0,0.75) {};
        \node[vtx] (v2) at (0,-0.75) {};
        \node[vtx] (v3) at (-0.75,-0.4) {};
        \node[vtx] (v4) at (0,0) {};
        \vtx{v0}
        \vtxIO{v1}
        \vtxIO{v2}
        \vtx{v3}
        \vtx{v4}
        \draw[edge] (v0) -- (v3);
        \draw[edge] (v0) -- (v4);
        \draw[edge] (v1) -- (v4);
        \draw[edge] (v2) -- (v4);
        \draw[edge] (v3) -- (v4);
        %
        %
        \node[tailv] (in0) at (-1,0.75) {};
        \node[tailv] at (-2,0.75) {};
        \node[tailv] (in1) at (-1,-0.75) {};
        \node[tailv] at (-2,-0.75) {};
        \node[tailv] (out0) at (1,0.75) {};
        \node[tailv] at (2,0.75) {};
        \node[tailv] (out1) at (1,-0.75) {};
        \node[tailv] at (2,-0.75) {};
        \node at (-2.5,0.75) {$\cdots$};
        \draw[tail] (-2.1,0.75) -- (in0);
        \node at (-2.5,-0.75) {$\cdots$};
        \draw[tail] (-2.1,-0.75) -- (in1);
        \node at (2.5,0.75) {$\cdots$};
        \draw[tail] (2.1,0.75) -- (out0);
        \node at (2.5,-0.75) {$\cdots$};
        \draw[tail] (2.1,-0.75) -- (out1);
        \draw[tail] (in0) .. controls +(0:0.5) and +(180:0.5) .. (v1);
        \draw[tail] (in1) .. controls +(0:0.5) and +(180:0.5) .. (v2);
        \draw[tail] (v1) .. controls +(0:0.5) and +(180:0.5) .. (out0);
        \draw[tail] (v2) .. controls +(0:0.3) and +(180:0.5) .. (out1);
      \end{tikzpicture}
    }\\
    \subfloat{\label{subfig:irrationalLength}
      \begin{tikzpicture}
        \node at (-4.25,0) {\subref{subfig:irrationalLength}};
        \node[vtx] (v0) at (-0.75,0.75) {};
        \node[vtx] (v1) at (-0.75,-0.75) {};
        \node[vtx] (v2) at (0.75,0.75) {};
        \node[vtx] (v3) at (0.75,-0.75) {};
        \node[vtx] (v4) at (-1.25,0) {};
        \node[vtx] (v5) at (0,1) {};
        \node[vtx] (v6) at (0,-1) {};
        \node[vtx] (v7) at (0,0) {};
        \vtxIn{v0}
        \vtxIn{v1}
        \vtxOut{v2}
        \vtxOut{v3}
        \vtx{v4}
        \vtx{v5}
        \vtx{v6}
        \vtx{v7}
        \draw[edge] (v0) -- (v4);
        \draw[edge] (v0) -- (v5);
        \draw[edge] (v0) -- (v7);
        \draw[edge] (v1) -- (v4);
        \draw[edge] (v1) -- (v6);
        \draw[edge] (v1) -- (v7);
        \draw[edge] (v2) -- (v5);
        \draw[edge] (v2) -- (v7);
        \draw[edge] (v3) -- (v6);
        \draw[edge] (v3) -- (v7);
        \draw[edge] (v5) -- (v7);
        \draw[edge] (v6) -- (v7);
        %
        %
        \node[tailv] (in0) at (-1.75,0.75) {};
        \node[tailv] at (-2.75,0.75) {};
        \node[tailv] (in1) at (-1.75,-0.75) {};
        \node[tailv] at (-2.75,-0.75) {};
        \node[tailv] (out0) at (1.75,0.75) {};
        \node[tailv] at (2.75,0.75) {};
        \node[tailv] (out1) at (1.75,-0.75) {};
        \node[tailv] at (2.75,-0.75) {};
        \node at (-3.5,0.75) {$\cdots$};
        \draw[tail] (-3.1,0.75) -- (in0);
        \node at (-3.5,-0.75) {$\cdots$};
        \draw[tail] (-3.1,-0.75) -- (in1);
        \node at (3.5,0.75) {$\cdots$};
        \draw[tail] (3.1,0.75) -- (out0);
        \node at (3.5,-0.75) {$\cdots$};
        \draw[tail] (3.1,-0.75) -- (out1);
        \draw[tail] (in0) .. controls +(0:0.5) and +(180:0.5) .. (v0);
        \draw[tail] (in1) .. controls +(0:0.5) and +(180:0.5) .. (v1);
        \draw[tail] (v2) .. controls +(0:0.5) and +(180:0.5) .. (out0);
        \draw[tail] (v3) .. controls +(0:0.3) and +(180:0.5) .. (out1);
      \end{tikzpicture}
    }\\
    \subfloat{\label{subfig:irrationalLengthIdentity}
      \begin{tikzpicture}
        \node at (-4.25,0) {\subref{subfig:irrationalLengthIdentity}};
        \node at (3.5,0) {\hphantom{$\cdots$}};
        \node[vtx] (v0) at (-0.5,0.75) {};
        \node[vtx] (v4) at (0.5,0.75) {};
        \node[vtx] (v5) at (-0.5,-0.75) {};
        \node[vtx] (v1) at (0.5,-0.75) {};
        \node[vtx] (v7) at (0,0) {};
        \node[vtx] (v6) at (-1,0) {};
        \node[vtx] (v2) at (-0.707,0.4) {};
        \node[vtx] (v3) at (-0.707,-0.4) {};
        \vtxIn{v0}
        \vtxOut{v4}
        \vtxIn{v5}
        \vtxOut{v1}
        \vtx{v7}
        \vtx{v6}
        \vtx{v2}
        \vtx{v3}
        \draw[edge] (v0) -- (v4);
        \draw[edge] (v0) -- (v7);
        \draw[edge] (v1) -- (v5);
        \draw[edge] (v1) -- (v7);
        \draw[edge] (v2) -- (v6);
        \draw[edge] (v2) -- (v7);
        \draw[edge] (v3) -- (v6);
        \draw[edge] (v3) -- (v7);
        \draw[edge] (v4) -- (v7);
        \draw[edge] (v5) -- (v7);
        \draw[edge] (v6) -- (v7);
        %
        %
        \node[tailv] (in0) at (-1.5,0.75) {};
        \node[tailv] at (-2.5,0.75) {};
        \node[tailv] (in1) at (-1.5,-0.75) {};
        \node[tailv] at (-2.5,-0.75) {};
        \node[tailv] (out0) at (1.5,0.75) {};
        \node[tailv] at (2.5,0.75) {};
        \node[tailv] (out1) at (1.5,-0.75) {};
        \node[tailv] at (2.5,-0.75) {};
        \node at (-3.25,0.75) {$\cdots$};
        \draw[tail] (-2.85,0.75) -- (in0);
        \node at (-3.25,-0.75) {$\cdots$};
        \draw[tail] (-2.85,-0.75) -- (in1);
        \node at (3.25,0.75) {$\cdots$};
        \draw[tail] (2.85,0.75) -- (out0);
        \node at (3.25,-0.75) {$\cdots$};
        \draw[tail] (2.85,-0.75) -- (out1);
        \draw[tail] (in0) -- (v0);
        \draw[tail] (in1) -- (v5);
        \draw[tail] (v4) -- (out0);
        \draw[tail] (v1) -- (out1);
      \end{tikzpicture}
    }
  \end{center}
  \caption{\label{fig:exampleGraphs}
    Some graphs that exhibit non-intuitive properties.
    At $k=\pi/3$, \subref{subfig:irrationalRotation} performs a $Z$ rotation
    through an angle of $\arctan\left(5\sqrt{3}/11\right)$, followed by a
    $Z$ gate.
    Also at $k=\pi/3$, \subref{subfig:fractionalLength} is an identity gate
    with an effective length of $\ell=1/2$.
    Finally at $k=\pi/4$, \subref{subfig:irrationalLength} performs the
    basis-changing
    operation $\sqrt{X}$ and \subref{subfig:irrationalLengthIdentity}
    implements an
    identity, each with the irrational effective length of $\ell=5-2\sqrt{2}$.
    (Supplemental material IDs 802, 488, 327, and 330.)
  }
\end{figure}
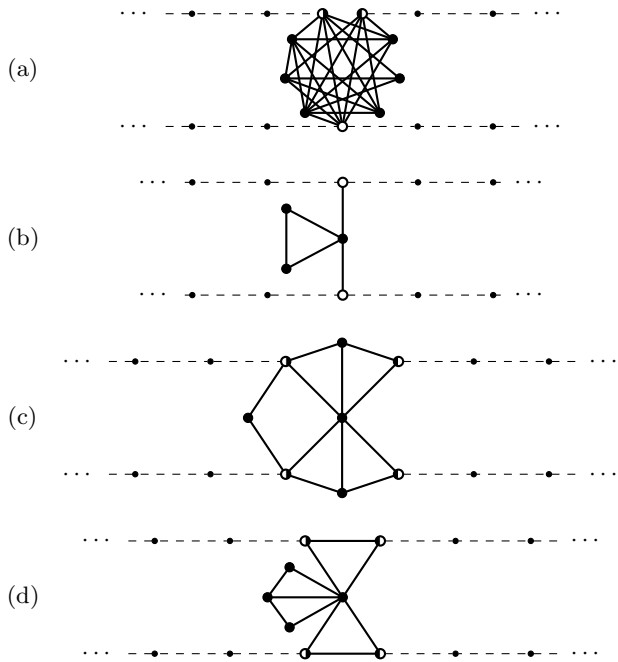

Figure~\ref{fig:exampleGraphs} showcases three graphs that produce
interesting and perhaps unexpected results.
While there is no reason to assume {\it a priori} that our procedure
will not find any rotations by irrational multiples of $\pi$, neither is
it intuitive that this should be the case given that the graphs are scattering 
plane waves with momentum values that are rational fractions of $\pi$.
Nevertheless, we indeed identify over 100 rotations through angles that
numerically appear to be irrational fractions of $\pi$.
These individual results
can be checked analytically by solving Eqs.~\eqref{eq:main}-\eqref{eq:transmit}
exactly in the cases of interest; those we checked bore out the apparent
irrationality indicated by the numerical results.
Figure~\ref{subfig:irrationalRotation} depicts one such case, a graph that
implements the transformation $Z R_Z[\arctan(5\sqrt{3}/11)]$ up to a global 
phase at momentum $k=\pi/3$.  Besides the novelty of obtaining irrational 
multiples of $\pi$ under the circumstances, these graphs are inherently useful 
because any two rotations about non-parallel axes by
irrational multiples of $\pi$ form a universal set for fixed-precision 
single-qubit quantum computation \cite{Kaye:2007}.

Similarly, it is not obvious whether any graph should have a non-integral
effective length.
Figure~\ref{subfig:fractionalLength} shows a graph on only five vertices that
is capable of acting as an identity gate with an effective length of
$\ell=1/2$;
at the momenta in question, no graph exists on fewer than
five vertices that has a non-integral effective length and implements a
single-qubit unitary.
Finally we see in Fig.~\ref{subfig:irrationalLength} a graph that 
has an irrational effective length of $\ell=5-2\sqrt{2}$. This graph implements
the basis-changing operation $\sqrt{X}$, and remarkably the graph in
Fig.~\ref{subfig:irrationalLengthIdentity} acts as an identity gate with this
same irrational length.
Of the 3118 graphs capable of implementing single-qubit unitaries with
commensurate identities, 2352 of them have non-integral effective lengths.
Of these, 1042 appear numerically to have irrational lengths, including
the `longest' graph identified, which has effective length
$\ell=350+156\sqrt{5}\approx698.826$ (ID 2174 in supplemental material
\cite{SupMat}).
This extreme effective length due to a comparatively small number of
vertices (i.e.\ $9\ll700$) is another observed phenomenon whose presence
is not initially obvious.
Such a length corresponds to the incoming wavepacket's having been
localized in the region of the graph for a significant duration, and is
reminiscent of a diverging negative scattering length, approaching unitarity
in traditional quantum scattering theory.
Almost 20\%\ of the unitaries identified have effective lengths $\ell\ge10$,
with greater than 1\%\ having $\ell\ge100$.
In the same vein, it is also noteworthy that no negative effective lengths
were identified.

\section{Conclusions}\label{sec:conclusions}
Inspired by the proof due to Childs that wavepackets scattering off small
simple graphs can perform universal quantum computation, we have exhaustively
studied scattering at nine momenta over the set of graphs on fewer than 10
vertices. As the number of vertices in the scattering center increases, so 
does the number of distinct single-qubit unitary operations that can be
performed.
The number of distinct graphs capable of producing these unitaries grows
super-exponentially, providing many methods by which to construct a given
operator.
The promise of investigating graphs on larger number of vertices is that
a desired operation might be able to be implemented by scattering off one
or a few graphs, where its decomposition into the two transformations in
a single universal set could require many more.

Rotations of the Bloch sphere can be performed about many axes distributed
roughly uniformly across its surface, by both rational and irrational
multiples of $\pi$. This results in a large number of distinct sets of
operators that are universal for quantum computation, which can be combined
together to minimize the total size of the graph used in a computation.
Natural extensions to this work include increasing the number of vertices
under investigation, the array of momenta studied, or both; and expanding
the number of tails attached to each scattering graph to find two- or
three-qubit gates (or more).

\section*{Acknowledgments}
We are grateful to Andrew Childs for helpful comments in the early stages of
this work. The authors acknowledge funding from Alberta Innovates -- Technology 
Futures (MSU), and the Natural Sciences and Engineering Research Council of
Canada.


\begin{thebibliography}{99}

\bibitem{Ambainis:2003}
A. Ambainis, Int. J. Quant. Inf.~{\bf 1}, 507 (2003).

\bibitem{Kempe:2003}
J. Kempe, Contemporary Physics~{\bf 44}, 307 (2003).

\bibitem{Kendon:2007}
V. Kendon, Math. Struct. Comp. Sci.~{\bf 17}, 1169 (2007).

\bibitem{Kendon:2011}
V. Kendon, e-print: arXiv:1107.3795 (2011).

\bibitem{Santha:2008}
M. Santha, in {\it Theory and Applications of Models of Computation}
(Springer Berlin / Heidelberg, 2008), vol. 4978, pp. 31-46.

\bibitem{Magniez:2005}
F. Magniez, M. Santha, and M. Szegedy, Proceedings of the sixteenth annual 
ACM-SIAM symposium on Discrete algorithms (SODA 2005), p. 1109 (2005).

\bibitem{Ambainis:2007}
A. Ambainis, SIAM Journal on Computing~{\bf 37}, 210 (2007).

\bibitem{Childs:2003}
A. M. Childs, R. Cleve, E. Deotto, E. Farhi, S. Gutmann, and D. A. Spielman,
Proceedings of the 35th ACM Symposium on Theory of Computing (STOC 2003), p. 
59 (2003).

\bibitem{Farhi:2008}
E. Farhi, J. Goldstone, S. Gutmann, Theory of Computing~{\bf 4}, 169 (2008).

\bibitem{Childs:2009a}
A. M. Childs, R. Cleve, S. P. Jordan, and D. Yeung,
Theory of Computing~{\bf 5}, 119 (2009).

\bibitem{Cleve:2008}
R. Cleve, D. Gavinsky, and D. L. Yonge-Mallo, in {\it Theory of Quantum 
Computation, Communication, and Cryptography} (Springer-Verlag Berlin, 
Heidelberg, 2008).

\bibitem{Reichardt:2008}
B. W. Reichardt and R. Spalek, Proceedings of the 40th ACM Symposium on Theory 
of Computing (STOC 2008), p. 103 (2008).

\bibitem{Ambainis:2010}
A. Ambainis, A. M. Childs, B. W. Reichardt, R. Spalek, and S. Zhang, SIAM J. 
Comput.~{\bf 39}, 2513 (2010).

\bibitem{Childs:2009b}
A. M. Childs, Phys. Rev. Lett.~{\bf 102}, 180501 (2009).

\bibitem{Lovett:2010}
N. B. Lovett, S. Cooper, M. Everitt, M. Trevers, and V. Kendon, Phys. Rev. 
A~{\bf 81}, 042330 (2010).

\bibitem{Underwood:2010}
M. S. Underwood and D. L. Feder, Phys. Rev. A~{\bf 82}, 042304 (2010).

\bibitem{Kaye:2007}
P. Kaye, R. Laflamme, and M. Mosca, {\it An Introduction to Quantum 
Computation}, pp. 70-71 (Oxford University Press, 2007).

\bibitem{SupMat}
See Supplemental Material included with this manuscript at
\url{http://www.arxiv.org}
for a complete list of unitary gates found.

\bibitem{Varbanov:2009}
M. Varbanov and T.~A. Brun, Phys. Rev. A~{\bf 80}, 52330 (2009).

\bibitem{McKay}
B.~D.~McKay, \url{http://cs.anu.edu.au/~bdm/nauty}.

\bibitem{GSL}
M.~Galassi et al., {\it GNU Scientific Library Reference Manual},
\url{http://www.gnu.org/software/gsl}.

\bibitem{LAPACK}
E.~Anderson et al., {\it LAPACK Users' Guide}
(Society for Industrial and Applied Mathematics, 1999).


\bibitem{Sloane:A000088}
N.~J.~A.~Sloane, {\it On-Line Encyclopedia of Integer Sequences},
\url{http://oeis.org/A000088}.


\end{thebibliography}
\end{document}